\documentclass[twocolumn,showpacs,aps,superscriptaddress,prd,notitlepage,showkeys,nofootinbib]{revtex4-1}

\usepackage{changes}

\usepackage[a4paper, total={7in, 8in}]{geometry}
\usepackage[normalem]{ulem}
\usepackage{multirow}
\usepackage{amssymb}
\usepackage{amsmath}
\usepackage{graphicx}
\usepackage{dcolumn}
\usepackage{hyperref}
\usepackage{parskip}
\usepackage{float}

\hypersetup{colorlinks,urlcolor=cyan,citecolor=cyan,linkcolor=cyan}
\usepackage{color,units}
\usepackage{lineno}
\usepackage{xspace}
\usepackage{longtable} 
\usepackage{float}  
\usepackage{amsfonts,wasysym,epsfig,verbatim,subfigure,bm,mathrsfs,lipsum}

\usepackage{color}
\usepackage{hyperref}
\usepackage{amsmath}
\usepackage{multirow}
\usepackage{graphicx}
\usepackage{envmath}
\usepackage{natbib}
\usepackage{lineno}

\definecolor{orange-red}{rgb}{1.0, 0.27, 0.0}

\newcommand{\IUCAA}{Inter-University Centre for Astronomy and
  Astrophysics, Post Bag 4, Ganeshkhind, Pune 411 007, India}

\newcommand{\WSU}{Department of Physics \& Astronomy, Washington State University,
1245 Webster, Pullman, WA 99164-2814, U.S.A}

\newcommand{\MPIHan}{Max-Planck-Institut f\"ur Gravitationsphysik (Albert Einstein Institute), Callinstr. 38, 30167 Hannover, Germany}

\newcommand{\LUHHan}{Leibniz Universität Hannover, 30167 Hannover, Germany}

\newcommand{\ozgrav}{Australian Research Council Centre of Excellence for Gravitational Wave Discovery (OzGrav), Australia}

\newcommand{\UWA}{Department of Physics, University of Western Australia, Crawley WA 6009, Australia}

\def\Sref#1{Sec.~\ref{#1}\xspace}
\def\Fref#1{Fig.~\ref{#1}\xspace}

\def\Eref#1{Eq.~(\ref{#1})\xspace}

\def\be{\begin{equation}}
\def\ee{\end{equation}}
\def\bea{\begin{eqnarray}}
\def\eea{\end{eqnarray}}
\def \D{{\cal D}}
\def \R{{\cal R}}
\def \G{{\cal G}}

\def \VG{{\mathcal V}_{\G}}
\def \Vp{{\mathcal V}_\perp}
\def \S{{\cal S}}
\def \P{{\cal P}}
\def \a{\alpha}
\def \b{\beta}
\def \M{{\mathcal M}}

\def \x{{\bf x}}
\def \y{{\bf y}}
\def \h{{\bf h}}
\def \n{{\bf n}}

\def \e{{\bf e}}
\def \v{{\bf v}}

\begin{document}

\setlength{\parindent}{1em}
\setlength{\parskip}{.5em}

\title[]
{Improved binary black hole searches through
better discrimination against
noise transients}

\author{Sunil Choudhary}
\email{sunil.choudhary@uwa.edu.au}
\affiliation{\IUCAA}
\affiliation{\ozgrav}
\affiliation{\UWA}

\author{Sukanta Bose}
\email{sukanta@wsu.edu}
\affiliation{\IUCAA}
\affiliation{\WSU}

\author{Sanjeev Dhurandhar}
\email{sanjeev@iucaa.in}
\affiliation{\IUCAA}

\author{Prasanna Joshi}
\email{prasanna.mohan.joshi@aei.mpg.de}
\affiliation{\MPIHan}
\affiliation{\LUHHan}

\begin{abstract}

Short-duration noise transients in LIGO and Virgo detectors significantly affect the search sensitivity of compact binary coalescence (CBC) signals, especially in the high mass region. In a previous work by the authors \cite{Joshi_2021}, a $\chi^2$ statistic was proposed to distinguish them, when modeled as sine-Gaussians, from non-spinning CBCs. The present work is an extension where we demonstrate the better noise-discrimination of an improved $\chi^2$ statistic -- called the optimized sine-Gaussian $\chi^2$ -- in real LIGO data. 
The extension includes accounting for the initial phase of the noise transients and use of a well-informed choice of sine-Gaussian basis vectors selected to discern how CBC signals and some of the most worrisome noise-transients project differently on them~\cite{sunil_2022}. 
To demonstrate this improvement, we use data with blip glitches from the third observational run (O3) of LIGO-Hanford and LIGO-Livingston detectors. 
Blips are a type of short-duration non-Gaussian noise disturbance known to adversely affect high-mass CBC searches.
For CBCs, 
spin-aligned binary black hole signals were simulated using the \textsc{IMRPhenomPv2} waveform and injected into real 
LIGO data from the same run. 
{We show that in comparison to the sine-Gaussian $\chi^2$, 
the optimized sine-Gaussian $\chi^2$ improves the overall true positive rate by around 6\% in a lower-mass bin ($m_1,m_2 \in [20,40]M_{\odot}$) and by more than 3\% in a higher-mass bin ($m_1,m_2 \in [60,80]M_{\odot}$). On the other hand, we see a larger improvement -- of more than 20\% -- in both mass bins in comparison to the traditional $\chi^2$.}
~

\end{abstract}

\maketitle

\section{Introduction}


Gravitational-wave (GW) astronomy has achieved several feats 
in recent years -- following up on  the first detection  of the binary black hole merger GW150914~\citep{GW150914}. After that first breakthrough detection, two LIGO
detectors (in Livingston and Hanford)~\cite{advligo} along with the Virgo detector (in Cascina)~\cite{AdvVIRGO} have observed more than 90 compact binary coalescence
(CBC) signals
from various kinds of binaries involving black holes (BHs) and neutron stars (NSs) in their first three observation runs \citep{gwtc3}. The fourth observation
(O4) run began in 2023, and is expected to include KAGRA~\citep{kagra}.
The GW community is expecting the CBC  detection rate to increase significantly in O4.
It is, therefore, important to find ways to effectively handle data
quality and detector characterization
to improve the search sensitivity so as 
not to miss interesting
signals. 
Currently,
high-mass CBC searches (for component mass $> 60 M_{\odot}$)
are adversely affected by noise transients~\cite{Nitz:2017lco, 2018_dq_vetoes} and 
some works 
have developed techniques to improve the search sensitivity in that part of the CBC parameter space~\cite{2017PhRvD..96j4015K,Nitz:2017lco,TanmayaMishra:2019,Joshi_2021, 2020CQGra..37n5001D, PhysRevD.104.064051,Ashton:2021tvz,sunil_2022}.
These works include
statistical, instrumental and, recently, a few machine-learning efforts. 

All studies about CBC search sensitivity in the high-mass region typically mention the impact of {\em blip} glitches~\cite{Cabero:2019orq} as a major source of deterioration. 
These  glitches are a type of short-duration non-Gaussian noise 
artifact
found in both LIGO detectors as well as in the Virgo detector. The duration of these glitches 
is around 10ms. In the frequency domain they are over 100Hz wide. Studies on the blips in O2 and O3~\cite{Cabero:2019orq,Davis_2021} 
mention that they occur 2-3 times per hour in both LIGO detectors. The reason behind blips affecting the CBC search sensitivity is that their time-frequency morphology has a lot of similarity with GW 
signals from 
CBCs with high total mass. 
These are essentially signals from binary black holes (BBHs). 

According to recent blip studies, their source is still not fully known~\cite{Cabero:2019orq}. 
These types of glitches do not show much correlation with any of the auxiliary channels (i.e., noise source monitoring channels). Therefore, it is tricky to confirm them as non-astrophysical in origin and remove them from 
short-duration signal searches.
One way to veto blips from GW data is to develop a statistical test that can differentiate 
them from CBC signals based on their different time-frequency characteristics, e.g., in spectrograms. There are $\chi^2$ statistics,
such as the traditional (or power) $\chi^2$~\cite{Allen2004} and sine-Gaussian $\chi^2$~\cite{Nitz:2017lco}, that are implemented in GW search pipelines to tackle  glitches, 
with, especially, the latter showing some success in discriminating against blips. 
There are yet other $\chi^2$s that check for signal consistency by employing expected SNR variation in time or across a CBC template-bank~\cite{chadthesis, cody_2017, sachdev_2019, chu2021spiir}. 
Still there remains room for improving  current blip discrimination methods.

In this work, we exploit the 
\emph{unified} $\chi^2$ formalism~\cite{Joshi_2021} to develop
a new $\chi^2$ statistic 
that incorporates information about how blip glitches and BBH signals project differently on a basis of sine-Gaussian functions.
Following that work, we call it the 
optimized sine-Gaussian  $\chi^2$ 
statistic.
We also tune it in real data
to specifically reduce the adverse impact of blip glitches 
on the sensitivity of spinning BBH searches.
Previous work on this $\chi^2$ statistic
in Ref.~\cite{Joshi_2021} had targeted simulated sine-Gaussian transients and non-spinning BBH signals.

This paper is organized as follows. In \Sref{framework}, we discuss  theoretical aspects of the optimized SG $\chi^2$, including a brief introduction to the general framework of 
$\chi^2$ statistics. \Sref{procedure} describes the procedure for constructing the optimized SG $\chi^2$. In particular, we  show how to identify the basis vectors for this statistic and how to employ singular-value decomposition to choose from them the most effective ones 
-- a limited few.  
\Sref{results} presents the results and performance of optimized SG $\chi^2$ in discriminating simulated spin-aligned BBH signals in real LIGO data from its third observation run (O3), which includes thousands of real blip glitches. Finally, in \Sref{discussion} we discuss the future applicability and prospects of this work.

\section{$\chi^2$ discriminators and their optimization}
\label{framework}

\subsection{General Framework}

The general framework for $\chi^2$ discriminators has been described in  Ref.~\cite{Dhurandhar_2017}. It shows how the various $\chi^2$ discriminators can be unified into a single discriminator, which can be appropriately termed as the
{\it unified} $\chi^2$.
In this framework, a data train $x(t)$ defined over a time interval $[0, T]$ is viewed as a vector $\x$. Such data trains form a vector space $\D$. Vectors in $\D$  will be denoted in boldface, namely, 
$\x, \y \in \D$. Since the detector strain
is typically sampled at a high rate, of ${\cal O}(10^3)\,$Hz, and the signals studied here can be as long as ${\cal O}(10^3)\,$sec, the data vectors can have large number of components, i.e., $N \sim 10^6$ or larger. Hence, $\D$ is essentially the $N$-dimensional real set $\R^N$. When additional structure is added to $\D$, namely, that of a scalar product, then it becomes a Hilbert space. 

Next consider the detector noise $n(t)$, which is a stochastic process defined over the time segment $[0, T]$. 
It has an ensemble mean of zero, and is stationary in the wide sense. A specific noise realisation is a vector $\n \in \D$, where $\n$ is in fact a random vector. Its one-sided power spectral density (PSD) is denoted by $S_n (f)$. If $\tilde{x}(f)$ and $\tilde{y}(f)$ are the Fourier representations of the vectors $\x$ and $\y$, respectively, then the scalar product of two vectors $\textbf{x}$ and $\textbf{y}$ in $\D$ is given by:
\begin{equation}
(\textbf{x},\textbf{y})=4\Re \int_{f_{\rm lower}}^{f_{\rm upper}} df \frac{\tilde{x}^{*}(f)\tilde{y}(f)}{S_h(f)} \,,
\label{inner_product}
\end{equation}
where the integration limits usually demarcate the signal band of interest, $[{f_{lower}},\,{f_{upper}}]$.
We have used an integral for the scalar product because the number of components of a data vector is very large, as argued above, and the continuum limit may be taken from a sum to an integral. 
\par

The $\chi^2$ discriminator is a mapping from $\D$ to positive real numbers and is defined so that its value for the signal is zero and for Gaussian noise has a $\chi^2$ distribution with a reasonable number of degrees of freedom, $p$. Typically, the number of degrees of freedom is a few tens to a hundred. 
If a template $\h$ is triggered, then the $\chi^2$ for $\h$ is defined by choosing a finite-dimensional subspace $\S$ of dimension $p$ that is orthogonal to $\h$, i.e., for any $\y \in \S$, we must have $(\y, \h) = 0$. Then the $\chi^2$ for the template $\h$ is defined as just the square of the $L_2$ norm of the data vector $\x$ projected onto $\S$. Specifically, we perform the following operations. Take a data vector $\x \in \D$ and decompose it as:
\be
\x = \x_{\S} + \x_{\S^\perp} \,,
\ee
where $\S^{\perp}$ is the orthogonal complement of $\S$ in $\D$. $\x_{\S}$ and $\x_{\S^\perp}$ are projections of $\x$ into the subspaces $\S$ and $\S^{\perp}$, respectively. We may write $\D$ as a direct sum of $\S$ and $\S^{\perp}$, that is, $\D = \S \oplus \S^{\perp}$.
\par
Then the required statistic $\chi^2$ is,
\be
\chi^{2} (\x) = \| \x_{\S} \|^2 \,.
\ee
The $\chi^2$ statistic so defined has the following properties. Given any orthonormal basis of $\S$, say $\e_{\a}$, with $\a = 1, 2, ..., p$ and $(\e_{\a}, \e_{\b}) = \delta_{\a \b}$, we obtain the following:

\begin{enumerate}

\item For a general data vector $\x \in \D$, we have:
\be
\chi^2 (\x) = \| \x_{\S} \|^2 = \sum_{\a = 1}^p |(\x, \e_{\a})|^2 \,. 
\ee   

\item Clearly, $\chi^2 (\h) = 0$ because the projection of $\h$ into the subspace $\S$ is zero, i.e., $\h_{\S} = 0$. 

\item Now, 
the noise $\n$ 
is taken to be stationary and Gaussian, with PSD $S_n(f)$ and mean zero. Therefore, the following is valid:
\be
\chi^2 (\n) = \| \n_{\S} \|^2 = \sum_{\a = 1}^p |(\n, \e_{\a})|^2 \,.
\ee
Observe that the random variables $(\n, \e_{\a})$ are independent and Gaussian, with mean zero and variance unity. This is because $\langle (\e_{\a}, \n) (\n, \e_{\b}) \rangle = (\e_{\a}, \e_{\b}) = \delta_{\a \b}$, where the angular brackets denote ensemble average (see \citep{Creighton:2011zz} for proof). Thus, $\chi^2 (\n) $ possesses a $\chi^2$ distribution with $p$ degrees of freedom. 
\end{enumerate}

For convenience, 
one is free to choose any {\em orthonormal} basis of $\S$. In an orthonormal basis  the statistic is manifestly $\chi^2$ since it can be written as a sum of squares of independent Gaussian random variables, with mean zero and variance unity.
\par

In the context of CBC searches, however, we have a family of waveforms that depend on several parameters, such as masses, spins and other kinematical parameters. We denote these parameters by $\lambda^a, ~~ a = 1, 2, ..., m$. The templates corresponding to these  waveforms are normalized, i.e., $\| \h (\lambda^a) \| = 1$. Then the templates trace out 
a manifold $\P$ -- the signal manifold --  which is a submanifold of $\D$. We now associate a $p$-dimensional subspace $\S$ orthogonal to the template $\h (\lambda^a)$ at each point of $\P$ -- we have a $p$-dimensional vector-space ``attached" to each point of $\P$. When done in a smooth manner, this construction produces a vector bundle with a $p$-dimensional vector space attached to each point of 
manifold $\P$.  We have, therefore, found a very general mathematical structure for the $\chi^2$ discriminator. Any given $\chi^2$ discriminator for a signal waveform $\h (\lambda^a)$ is the 
{square of the} $L_2$ norm of a given data vector $\x$ projected onto the subspace $\S$ at $\h (\lambda^a)$. 
\par

It can be easily shown that the traditional  $\chi^2$ falls under the class of \emph{unified} $\chi^2$. This is done by exhibiting the subspaces $\S$ or by exhibiting the basis vector field for $\S$ over $\P$; the conditions mentioned above must be satisfied by $\S$. In \citep{DGGB2017} such a basis field has been  exhibited explicitly.

\subsection{Optimizing the $\chi^2$ discriminator}
\label{Opt_chi}

The $\chi^2$ discriminator must produce as large a value as possible for a glitch in the data. In our framework we achieve this, on average, given the collection of glitches. The optimization is therefore carried out for a family of glitches, say, $\G$. Here we will model the glitches as sine-Gaussians and select a family of such glitches based on the ranges of the parameters describing the sine-Gaussians. The subspace $\S$ then must be chosen in such a way as to have maximum projection on an average. Also one must keep in mind that $\S$ must be orthogonal to the trigger template. These two criteria essentially guide us to obtain the subspaces $\S$. The third criterion is that its dimension should be kept small in order to keep the computational cost at a reasonable level.    
\par

More specifically for a given trigger template $\h$, we perform the following steps:
\begin{enumerate}
 \item Sample the parameter space $\G$ of the glitches (sine-Gaussians) sufficiently densely so that the sample is representative. We call the subspace of $\D$ spanned by these sampled vectors as $\VG$. 
 This is done efficiently and conveniently with the help of a metric, as will be described in Sec.~\ref{select}. 
 
 \item Since $\S$ should be orthogonal to $\h$, we remove the component parallel to $\h$ from each of the sample vectors spanning $\VG$. Thus if $\v \in \VG$, then we define $\v_\perp = \v - (\v, \h) \h$. These vectors $\v_\perp$ by construction are orthogonal to $\h$. The space spanned by these clipped
 vectors $\v_\perp$ is called $\Vp$.  
 
 \item Next we apply Singular Value Decomposition (SVD) to the row vectors of $\Vp$ to obtain the best possible approximation of lower dimension say $p$. We will put a cut-off on the singular values so that projection obtained is as large as desired, say, $90\%$. The singular vectors corresponding to the singular values obtained by applying the cut-off generate the subspace $\S$. 
\end{enumerate}

Steps 1 and 2 above were also described in the construction of $\chi^2$ tests in Ref.~\cite{HF11} where the vectors $\v$ were taken to be {\it gravitational waveforms}. In our formulation, however, the sampled vectors in principle could be {\it any} vectors in $\D$, the only condition being that they be orthogonal to $\h$. Further, in order to construct an effective $\chi^2$, {\it the vital step is to choose these vectors in the direction of the glitches}. This is distinct and different from the waveforms suggested in  Ref.~\cite{HF11}. For more details on our formulation, we refer to Ref.~\citep{DGGB2017}, where we also proved that statistically independent $\chi^2$ discriminators, such as the traditional one and the optimized SG $\chi^2$, can be straightforwardly added to form new discriminators with a larger number of degrees of freedom that continue to have the $\chi^2$ distribution in Gaussian noise. 
{Such discriminators will tend to be effective against a broader class of glitches.}
In the next section, Sec.~\ref{select},  we describe how one can 
define a metric on $\G$
to obtain the sample vectors that 
span $\VG$. 

\section{Constructing the optimized sine-Gaussian $\chi^2$ discriminator}
\label{procedure}

\subsection{Selection of vectors that span $\VG$}
\label{select}

In order to form an optimal $\chi^2$, it is important that we select appropriate vectors in $\VG$
to project the GW data on. In order to improve the sensitivity of CBC searches, we specifically target the blip glitches in this work, which are a major source of 
reduction
in CBC search sensitivity. A recent work  \cite{Joshi_2021} demonstrates how transient bursts represented by sine-Gaussian waveform \cite{chatterji_thesis} can be vetoed with the help of an optimal $\chi^2$ from the GW data. 
Since the blip glitches are also known to have a time-domain morphology similar to the sine-Gaussian waveforms, we use these waveforms to form the vectors in $\VG$ for constructing the optimized SG $\chi^2$ here.
To allow for an arbitrary  phase in the noise transient, we use a complex-valued sine-Gaussian waveform. This is an improvement over our earlier work~\cite{Joshi_2021}, especially relevant
in dealing with the general scenario.
In the time domain the sine-Gaussian waveform can be defined as,
\begin{equation}
\begin{aligned}
\psi(t;t_0,f_0,Q) =   & A\exp{\bigg(-\frac{4\pi^2f_0^2}{Q^2}(t-t_0)^2\bigg)} \\
& \exp{[-i2\pi f_0(t-t_0)]} \,. 
\end{aligned}
\end{equation}
In the frequency domain, it is
\begin{equation}
\begin{aligned}
\tilde{\psi}(f;t_0,f_0,Q) = & \tilde{A}\exp{\bigg(-\frac{Q^2}{2\pi f_0^2}(f-f_0)^2\bigg)} \\
&\exp{[-i 2 \pi t_0(f-f_0)]}\,,
\end{aligned}
\end{equation}
%
where
$t_0$ is central time, $f_0$ is central frequency, 
$Q$ is the
quality factor, and the amplitudes are $A=\bigg(\frac{8\pi f_0^2}{Q^2}\bigg)^{\frac{1}{4}}$ and $\tilde{A}=\bigg(\frac{Q^2}{2\pi f_0^2}\bigg)^{\frac{1}{4}}$. 
{It is important to note that this model for the glitches is different from the one considered in \cite{Joshi_2021} because this model considers sine-Gaussians with {\it arbitrary} phase, which is more general. Therefore, the results that follow here are markedly different, and the resulting discriminator is effective more generally.} 

To calculate the metric in the ($t_0,f_0,Q$) space we begin by considering two
neighboring
sine-Gaussian waveforms in that space, namely,
$\psi_1(f;t_0,f_0,Q)$ and
$\psi_2(f;t_0+dt_0, f_0+d f_0,Q+d Q)$. A metric may then  be introduced on this space as %
a map from the differences in the parameters of these waveforms to the fractional change in their match:
%
\begin{equation}
\begin{aligned}
ds^2 = & \bigg(\frac{4 \pi^2 f_0^2}{Q^2}\bigg)dt_0^2+\bigg(\frac{2+Q^2}{4f_0^2}\bigg)df_0^2+\bigg(\frac{1}{2Q^2}\bigg)dQ^2 \\
       & -\bigg(\frac{1}{2 f_0 Q}\bigg)df_0 dQ. 
\end{aligned}
\label{metric}
\end{equation}
We do not consider the noise power spectral density (PSD)
to calculate the metric 
since it has a negligible effect on the arrangement of 
the vectors in $\VG$~\cite{chatterji_thesis}. The above metric (in \Eref{metric}) can be reduced to its diagonal form using the transformations,
\begin{equation}\omega_o=2\pi f_0
\end{equation}
and 
\begin{equation}
\nu=\frac{\omega_o}{Q}\,.
\end{equation}
In the new coordinates, $(t_0,\omega_0,Q)$, the  metric takes the form
\begin{equation}
    ds^2=\nu^2 dt_0^2 + \frac{1}{4\nu^2} d\omega_0^2 + \frac{1}{2 \nu^2} d\nu^2.
    \label{intmid_metric}
\end{equation}
In comparison to the metric in Ref. \cite{Joshi_2021}, this metric has no $\omega_0$ term multiplying $dt_0^2$. This results from our accounting for the aforementioned arbitrary {\em phase} of the sine-Gaussian waveform in this work.

As mentioned in Ref. \cite{Bose_2016}, a CBC template is triggered with a time lag $t_d$ after the occurrence of a glitch, 
i.e., after $t_0$.
The time $t_d$ is given by \cite{Bose_2016},

\begin{equation}
    t_d \simeq \tau_0 \bigg(1-\frac{16}{3Q^2}\bigg(\zeta+\frac{2}{3}\bigg)\bigg)\,,
\label{chirptime}    
\end{equation}
where the second term inside the parentheses determines the magnitude of the ``correction" beyond the chirp time $\tau_0$, which is given by
\begin{equation}
    \tau_0=\frac{5}{256\pi f_0}(\pi \mathcal{M}f_0)^{-5/3},
\end{equation}
and $\zeta$ is negative of the logarithmic derivative of the noise PSD ($S_h(f)$) evaluated at $f_0$.

The metric in \Eref{intmid_metric} can be reduced to a more simplified form with the help of transformations 
\begin{equation}
z=(\omega_0 \mathcal{M})^{-5/3}\quad {\rm and} \quad
y=\ln (\nu/\rm{rad/sec})\,,   
\end{equation}
such that
\begin{equation}
ds^2=\Bigg[\frac{2^{-14/3}}{Q^2}+\frac{9Q^2}{100 z^2} \Bigg]dz^2+\frac{1}{2}dy^2.
\label{final_metric}
\end{equation}
By examining Eq.~(\ref{chirptime}), one may reckon that for  $Q \sim 2$ the $t_d$ would be significantly affected. This is in fact so but it turns out that it makes little difference to the metric. This can be seen  as follows. In Eq.~(\ref{chirptime}) the correction term is inversely proportional to $Q^2$; therefore, for high $Q$, say, with $Q \gtrsim 5$, it is negligible. 
Also, at high central frequencies, i.e., $f_0 \gtrsim 500$ Hz, the factor $(\zeta + 2/3)$ is small for the aLIGO design
PSD. Thus, in either case one has $t_d \approx \tau_0$. 
The remaining case is one where $f_0 \sim 100$ Hz and $Q \sim 2$. Now the term arising from $t_d$ is the second term multiplying $dz^2$ in the metric expression~\Eref{final_metric}. It turns out that this term is small compared to the first one in the same metric expression -- about 14\% at this frequency. Hence, even if the $t_d$ is changed significantly, it makes a small difference to the metric. To summarize, in most of the parameter space we consider, the metric given in Eq.~(\ref{final_metric}) applies and our sampling based on this metric valid to a good approximation.
\par

From a broader perspective, the main idea is to sample the parameter space of glitches adequately, so that they are not misidentified. 
Thus, any inadequacy resulting from inaccuracies of the metric can be easily remedied by more densely sampling the parameter space. This can be achieved by just increasing the match between neighbouring points. We have explored this possibility and find that it does not lead to dissimilar results.

\begin{figure*}

\includegraphics[scale=0.85]{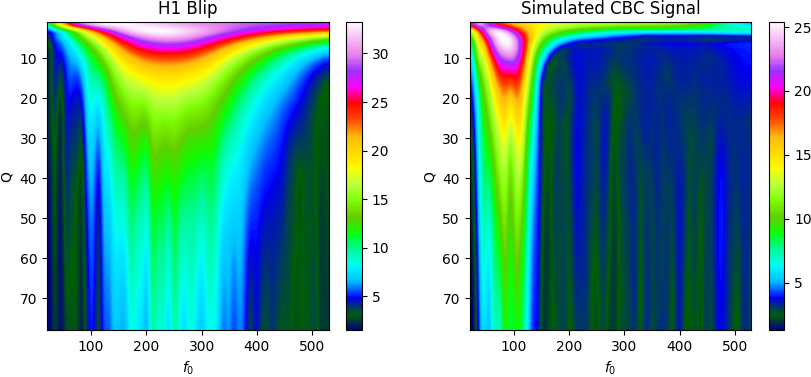}

\caption{The sine-Gaussian projection map of a blip with SNR = 16 (left) from the O3 LIGO-Hanford data and a simulated BBH signal with the same SNR (right)
injected in real noise from adjacent quieter data. Here the SNRs for the blip and the CBC signal are obtained via identical matched-filtering computations using the same BBH template-bank. The colorbar represents relative strength of the projections~\citep{sunil_2022}. 
{Note, however, that there is a variation among blips regarding how their power projects on sine-Gaussians across $f_0$. How well they can be discriminated from BBH signals in any mass range, therefore, is best studied for a wide population of blips as pursued below.}
} 
\label{SGP_map}
\end{figure*}

\begin{figure*}
\includegraphics[scale=0.5]{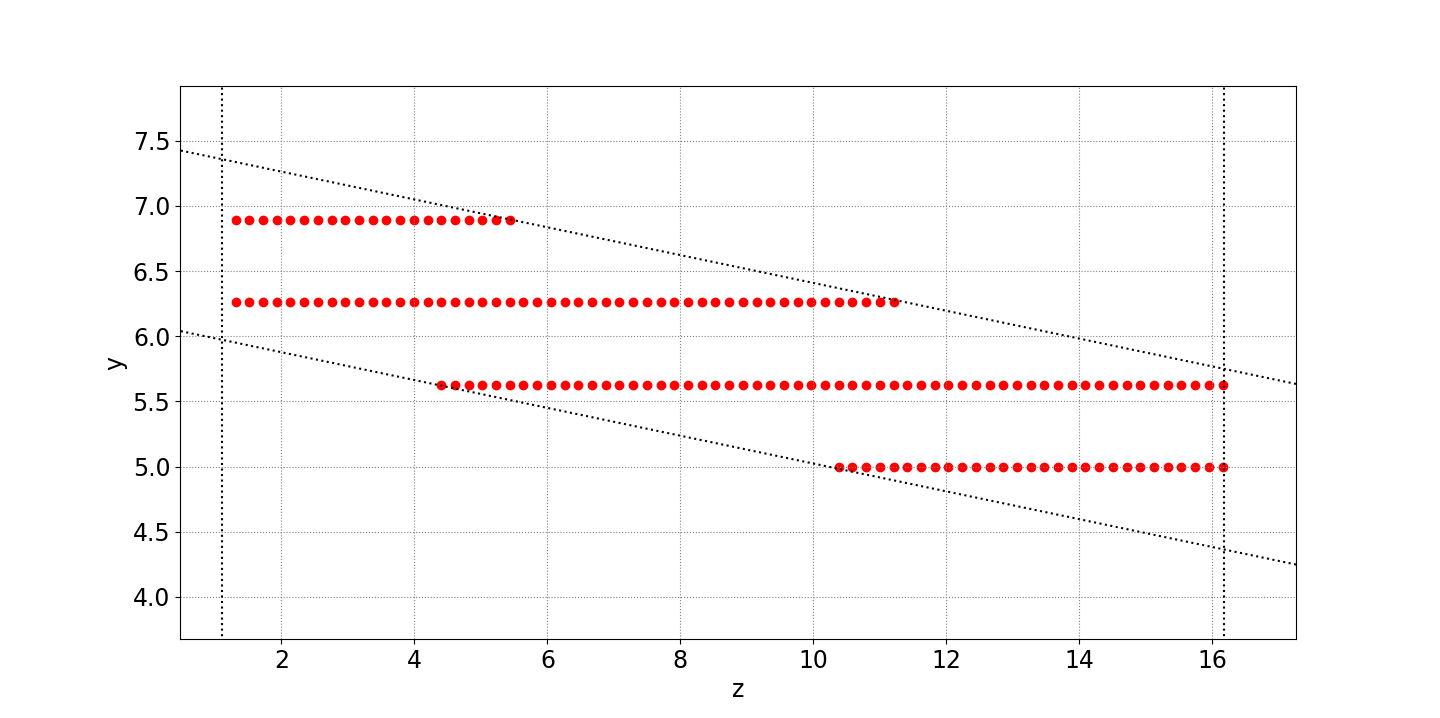}
\includegraphics[scale=0.5]{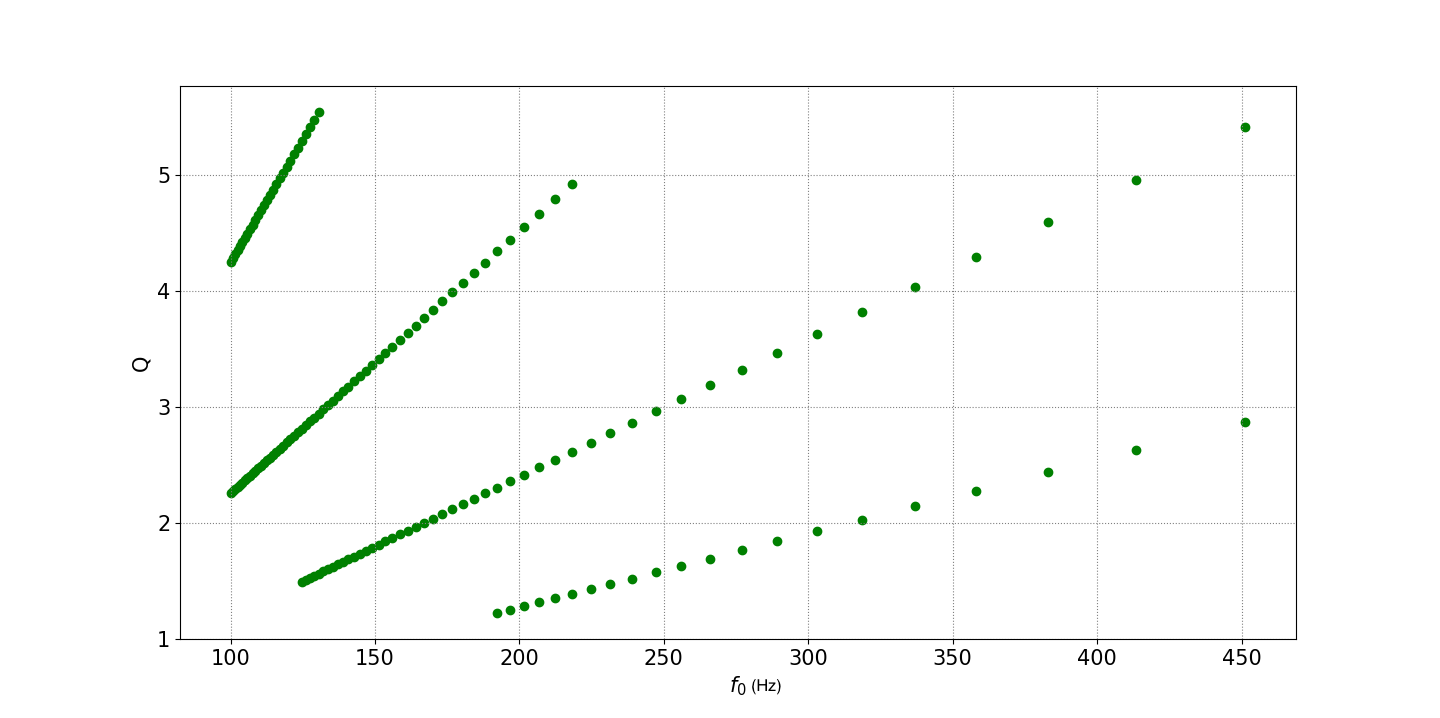}
\caption{
Sampled points in the $z-y$ space (top) for a triggered template with component masses (70,70)~$M_{\odot}$, and a  chirp mass of 60.9~$M_{\odot}$. The rectangular region $f_0\in[100,500]\,$Hz and $Q\in[2,8]$, in the $f_0-Q$ space (bottom), transforms to a trapezium in the $z-y$ space.
Neighboring points have a minimum match of 80\% 
and are essentially uniformly placed in the $z-y$ plane. There are a total of 58 
sampled points in this plot. This number can vary with the chirp mass of the triggered template.}
\label{z_y}
\end{figure*}


To choose the appropriate vectors in $\VG$ for the optimal-$\chi^2$, we take help from the sine-Gaussian projection maps introduced in Ref. \cite{sunil_2022}. The SGP maps are a projection of GW data onto the sine-Gaussian parameter space. We experimented with several real blips and simulated CBC signals to model their projections on the sine-Gaussian parameter space. As 
{seen in \Fref{SGP_map}, blips and CBC signals show} projection in distinct regions of the sine-Gaussian parameter space. Blips 
project strongly in the frequency region above 100 Hz, whereas for the CBC signals with component masses above 10~$M_{\odot}$ the projection 
{lies} mostly below 100~Hz. Along the $Q$ coordinate, the CBC signals show more elongated features than the blips. This difference in the projections of blips and CBCs on sine-Gaussians
paves the way for selecting appropriate vectors in $\VG$ to formulate an optimal-$\chi^2$ following the \emph{unified} $\chi^2$ formalism \cite{Dhurandhar_2017}.
We choose parameter ranges such that the blips have a high projection on these vectors that
lead to higher values of the $\chi^2$ statistic 
for blips 
than CBC signals. In our case, we find $f_0\in[100,500]$~Hz and $Q\in[2,8]$. 

To construct the $\VG$  vectors in the chosen region of the sine-Gaussian parameter space we first use \Eref{final_metric} to sample points in the $z$-$y$ space. The coefficients of $dz^2$ depend on $Q$, $f_0$ and the chirpmass $\mathcal{M}$ through the parameter $z$. To get a flat metric in $z$-$y$ space, we fix $Q=8$ and $f_0=500$ Hz. The choice of $Q$ and $f_0$ is made after observing that it provides sufficiently dense sampled 
points such that the mismatch between two neighbouring 
vectors is not more than 0.20. After sampling 
points in the $z$-$y$ space, we transform them back to the $Q$-$f_0$ space. 
{The top panel of \Fref{z_y} shows 
points sampled in the $z$-$y$ space, while the bottom panel shows the same points after transforming to the $Q$-$f_0$ space.} Sine-Gaussian waveforms are chosen corresponding to each of these sampled points. The number of total 
points can vary depending on the chirp mass $\mathcal{M}$ of the triggered template. This can lead to a large number of 
vectors. 
To reduce that number, note that most of them are not linearly independent.
We, therefore, use singular value decomposition (SVD) 
(discussed below in \Sref{SVD})
to obtain the (much smaller number of) basis vectors from that sample.
The resultant SVD vectors are then used to project the GW data upon them using \Eref{inner_product}. As it turns out, we require 
{just  about} 
three vectors -- on which the data show maximum projection -- to compute the optimal-$\chi^2$, which is then defined as,
\begin{equation}
    \chi_{\rm opt}^2=\sum_{\alpha=1}^{3}|(\textbf{x},\textbf{g}_{\alpha})|^2 \,,
\label{eq:chisqbasis}
\end{equation}
where $\textbf{x}$ is the data vector and $\textbf{g}_{\alpha}$ are the three aforementioned  
basis vectors. 


\subsection{Reducing the number of degrees of freedom of the $\chi^2$}
\label{SVD}

The selected vectors (sine-Gaussians) as described in Section \ref{select} usually turn out to be quite large in number. After time-shifting the sine-Gaussians and clipping their components parallel to the template, they span the subspace $\Vp$ (see Section \ref{Opt_chi}). We could in principle use $\Vp$ on which to project the data vector and compute the $\chi^2$ statistic. But the dimension of $\Vp$ is usually large and  
in practice it would involve too much computational effort and slow down the search pipeline -- the $\chi^2$ would involve too many degrees of freedom, namely, the dimension of $\Vp$. Therefore, we look for the best $p$-dimensional approximation to $\Vp$, where $p$ is reasonably small. The SVD algorithm allows us to achieve just this -- this is the essence of the Eckart-Young-Mirsky theorem \citep{EY1936}.
\par

However, the SVD cannot be applied directly to the selected vectors in $\VG$. The input matrix, say, $M$ needs to be first prepared. We briefly describe the procedure here since the full details can be found in \citep{Joshi_2021}. The following are the salient steps needed:

\begin{itemize} 
\item The sine-Gaussians have central time $t_0 = 0$ and they need to be appropriately time-shifted with respect to the time of occurrence of the trigger. We will always take the trigger to occur at $t = 0$, and so the glitch must have occurred at time $- t_d$. Accordingly the sine-Gaussians have to be shifted by the time $-t_d$. Since we write the input matrix in the Fourier domain, this is achieved by multiplying each row vector by the phase factor 
$e^{2 \pi i f t_d}$.

\item  The selected vectors need to be {\em clipped} by subtracting out from each sine-Gaussian its component parallel to the template. The clipped and time-shifted sine-Gaussians span the subspace $\Vp$ of $\D$. The desired subspace $\S$ is a subspace of $\Vp$.

\item The usual SVD algorithm ``sees" the Euclidean scalar product. However, here we have a weighted scalar product -- inversely weighted by the PSD $S_h (f)$. So in order to apply the usual SVD algorithm we need to {\it whiten} each row vector. This is achieved by dividing each Fourier component of the row vector by $\sqrt{S_h (f)}$.

\item The input matrix is now ready to be fed into the SVD algorithm.
\end{itemize}

This is however not the end of the story. We have to also modify the output - which are the right singular vectors. We need to {\it unwhiten} these vectors by multiplying by the factor $\sqrt{S_h (f)}$. The unwhitened singular vectors are orthonormal in the weighted scalar product.
\par

The SVD algorithm also yields singular values $\sigma_i, ~i = 1, 2, ...,r$. The Frobenius norm of the input matrix is just $\parallel M \parallel_F = \sum_{i = 1}^r \sigma_i^2$. The number of degrees of freedom $p$ are chosen so that:
\be
\sum_{i = 1}^p \sigma_i^2 ~\gtrsim ~(1 - \delta)~\parallel M \parallel_F \,, ~~~~~~~~~~~~p ~\leq ~r ,
\ee
where $\delta$ may be chosen to be, say, $0.1$ or 10 \%. The subspace $\S$ is generated by the first $p$ right singular vectors and so has dimension $p$. This also means that we have a projection of about $90\%$, on the subspace $\S$.  $\S$ is the best $p$ dimensional approximation to $\Vp$ -- it is essentially a $p$-dimensional least-square fit to $\Vp$ (see \citep{Joshi_2021} for more discussion). We have also succeeded in  reducing the number of degrees of freedom of the $\chi^2$ to $p$.  Typically, for the ranges of parameters $f_0, ~Q, ~\M$, etc., considered here, the number of selected vectors in $\VG$ is about a few hundred
while $p < 10$.  Thus, $p$ is much smaller than the number of selected vectors in $\VG$.

\section{Results}
\label{results}

\subsection{Performance on Blips}

In this study, we tuned the optimized SG $\chi^2$ specifically for the blip glitches. In order to test the effectiveness of that $\chi^2$ in differentiating blips from {\em aligned spin} BBH signals, we chose real blips from LIGO's O3 run, as identified by the Gravity Spy tool \cite{Gravityspy1,Gravityspy2,Gravityspy3,Gravityspy4, Gravityspy5}. Blips were selected from both Hanford and Livingston detectors when they had 
a confidence level of 0.6 on 
a scale of 0 to 1, as rated by Gravity Spy. 
A total of 4000 strain data segments, each of length 16 sec and containing a blip with a matched-filtering signal-to-noise ratio (SNR) between {4 and 12}, as registered by the loudest BBH template, were chosen. The BBH data sample is prepared by simulating  BBH signals 
using the family of \textsc{IMRPhenomPv2} waveforms ~\cite{PhenomP}.  These simulated  signals also span the same SNR range, namely, 4 to 12, uniformly. We first divide the signals into two bins based on their component masses: One bin consists of signals with component masses $m_1,m_2 \in [20,40] M_{\odot}$ and the other consists of those with $m_1, m_2 \in [60,80] M_{\odot}$. In both cases, the aligned spins $s_{1z}$ and $s_{2z}$ are distributed uniformly in the range $[0.0, 0.9]$. The purpose of this division is to 
compare the performance of the optimal SG $\chi^2$ in the two mass bins and use it to understand which part of the BBH parameter space 
benefits more from its use in reducing the adverse impact of the blips. 
For the studies in this section, we needed 3 (complex) basis vectors in Eq.~(\ref{eq:chisqbasis}) for the construction of the optimized SG $\chi^2$, which amounts to 6 degrees of freedom for that statistic.

\begin{figure*}
\includegraphics[scale=0.28]{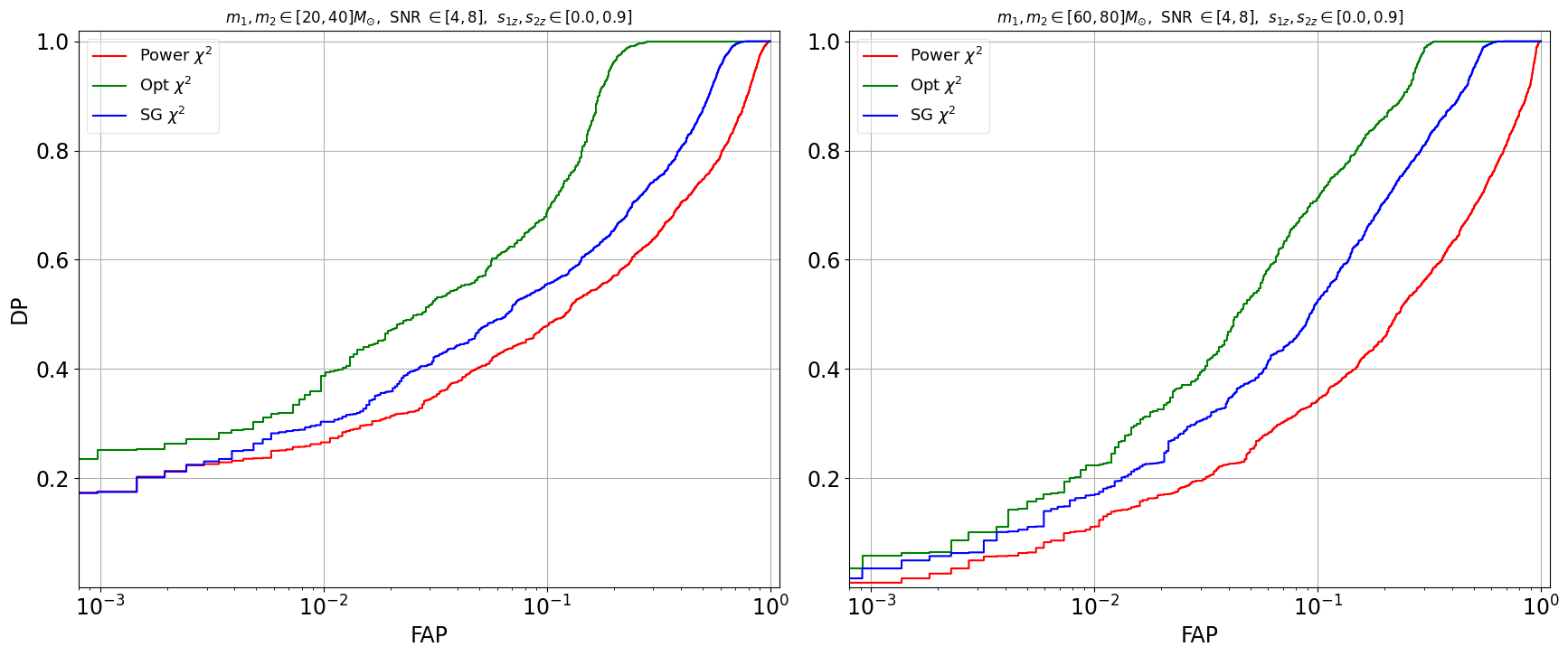}
\caption{The ROC plots for blips vs BBH signals for lower (left) and higher (right) mass bins.  Here, we use 2000 blips from real LIGO-Hanford O3 data and 2000 simulated CBC signals with component masses uniformly distributed in range $[20,40] M_{\odot}$ (left) and $[60,80] M_{\odot}$ (right). The aligned spins are in the range $[0.0, 0.9]$. To calculate the SNR in this case we have used a CBC template bank with component masses in $(m_1,m_2)_{bank} \in [10, 50] M_{\odot}$ and $(m_1,m_2)_{bank} \in [50, 90] M_{\odot}$, respectively. The SNRs are kept in the range [4,8].}
\label{scatter_1}
\end{figure*}

\begin{figure*}
\includegraphics[scale=0.28]{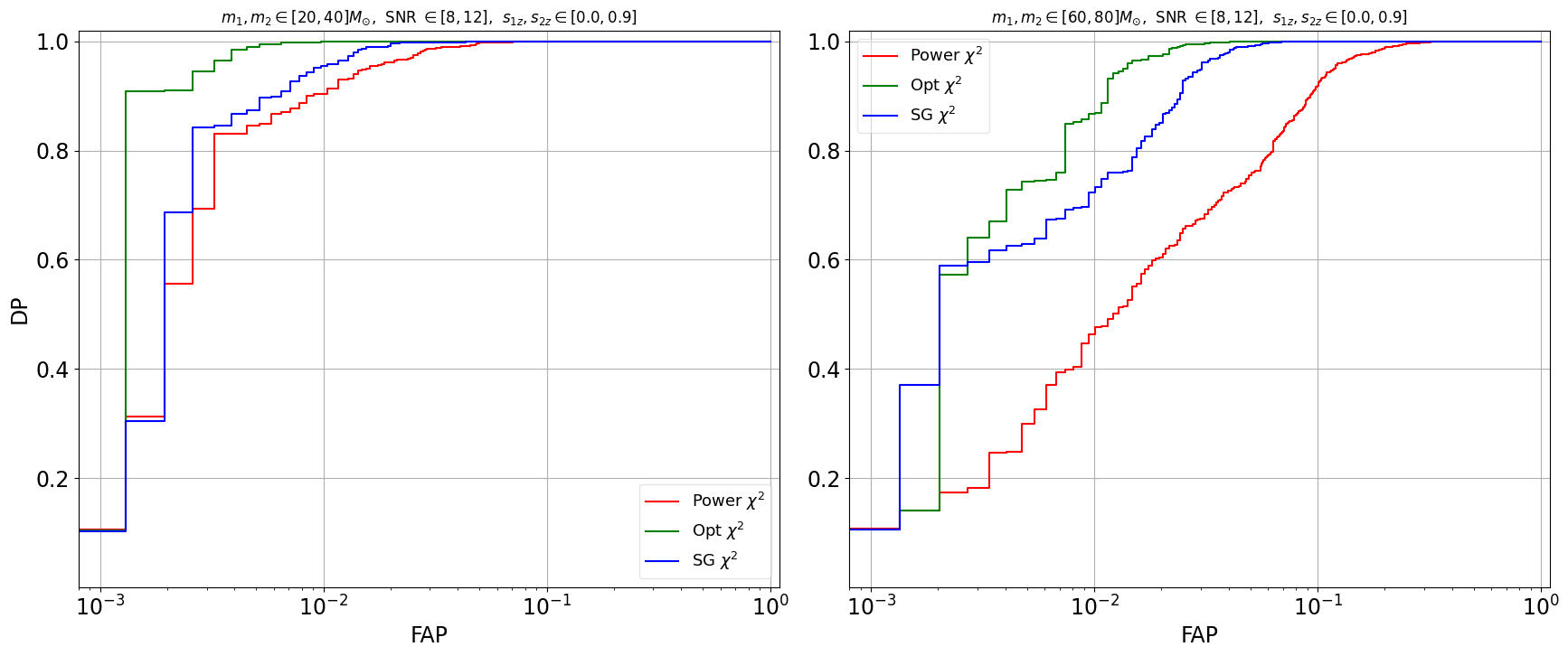}
\caption{This figure is similar to Fig.~\ref{scatter_1} but for the higher SNR range of [8,12]. Here too we use 2000 blips 
from real LIGO-Hanford O3 data
and 2000 simulated CBC signals with component masses uniformly distributed in range $[20,40] M_{\odot}$ (left) and $[60,80] M_{\odot}$ (right). The aligned spins are in the range $[0.0, 0.9]$. To calculate the SNR in this case we used a CBC template bank with component masses $(m_1,m_2)_{bank} \in [10, 50] M_{\odot}$ and $(m_1,m_2)_{bank} \in [50, 90] M_{\odot}$, respectively. } 
\label{scatter_2}
\end{figure*}

\begin{figure*}
\includegraphics[scale=0.28]{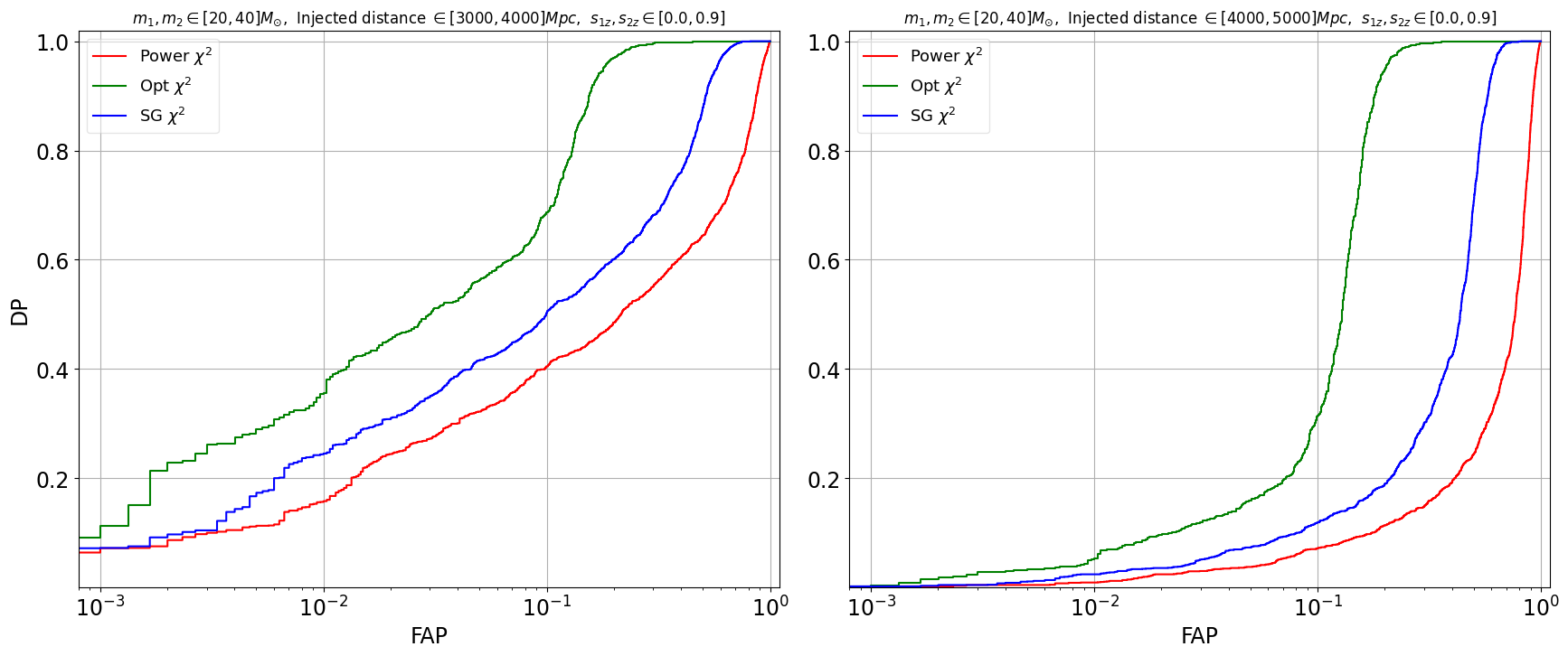}
\includegraphics[scale=0.28]{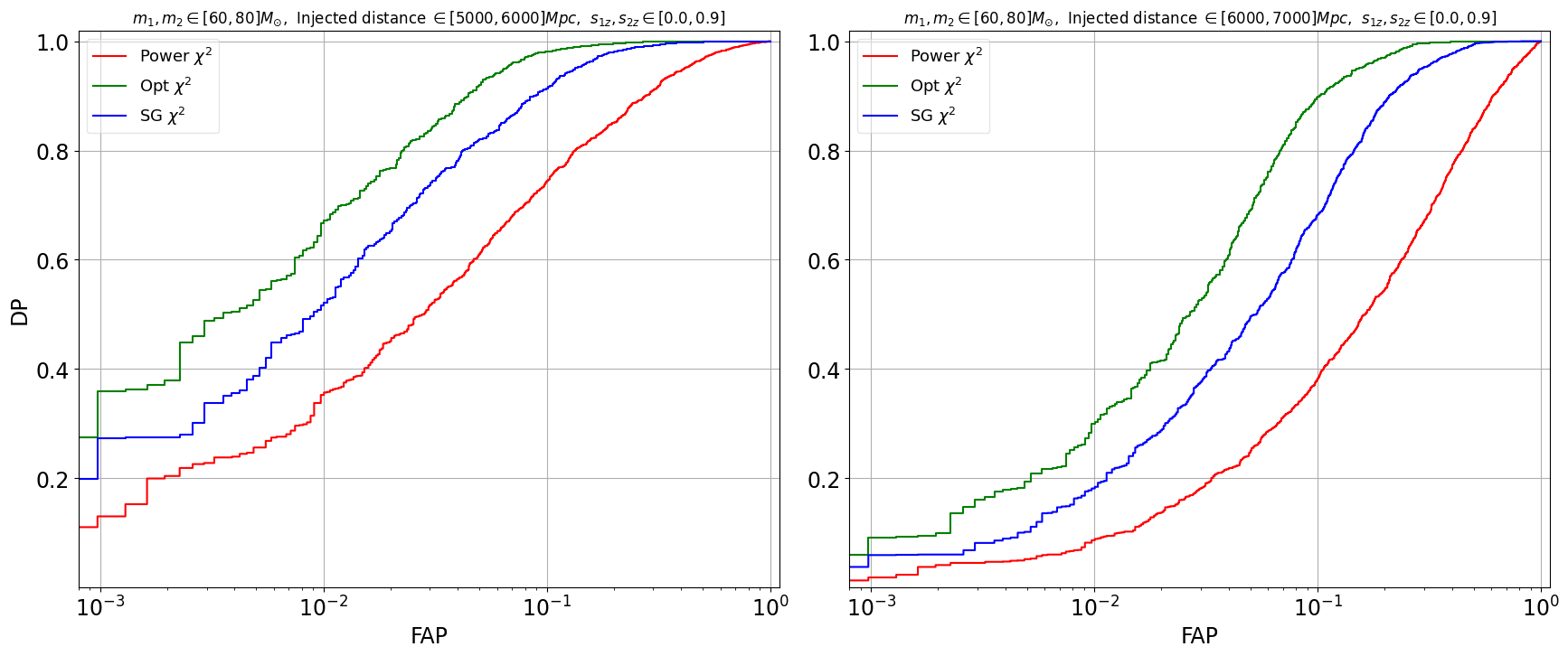}

\caption{Comparison of the three $\chi^2s$ for different source distance bins. For $m_1,\>m_2\in [20,40] M_{\odot}$ we have divided the source distance into two bins, namely, [3000,\>4000]~Mpc and [4000,\>5000]~Mpc. For  $m_1,\>m_2\in [60,80] M_{\odot}$, the source distance is divided into distance bins [5000,\>6000]~Mpc and [6000,\>7000]~Mpc. The blips chosen for analysis are in the SNR range $[4,10]$. We have specially chosen low-SNR blips since these are more difficult to distinguish from BBH signals.
}
\label{dist_1}
\end{figure*}

\begin{figure}
\includegraphics[width=\columnwidth]{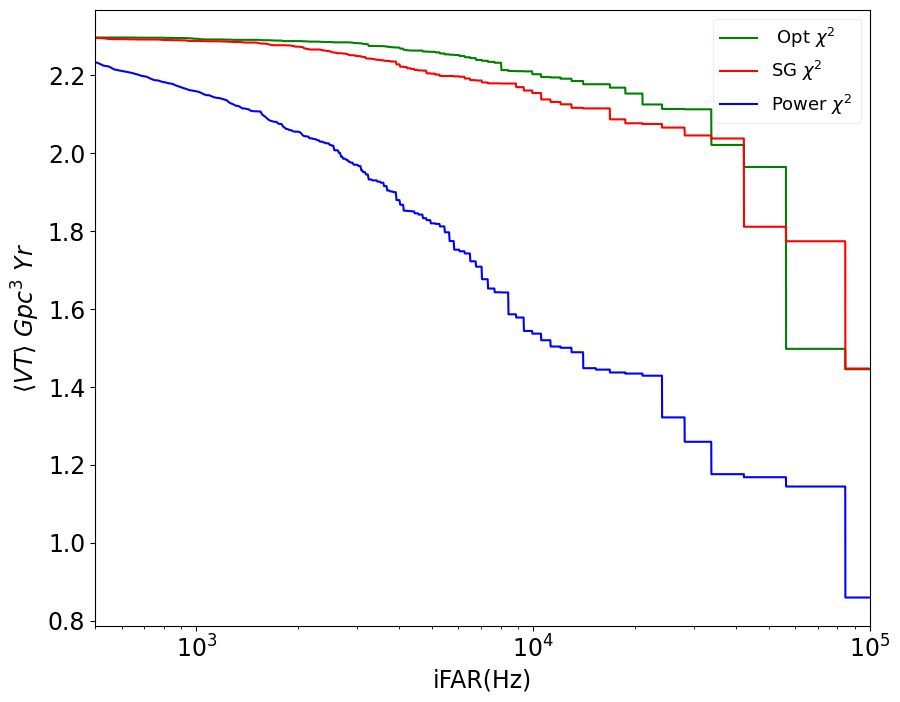}

\caption{ Volume-time sensitivity (VT) {\it vs} the inverse false-alarm rate (iFAR) compared for three different  $\chi^2$ statistics in the higher mass bin of $m_1,m_2\in [60,80]M_\odot$. 
}
\label{vt_1}
\end{figure}

\begin{figure}

\includegraphics[width=\columnwidth]{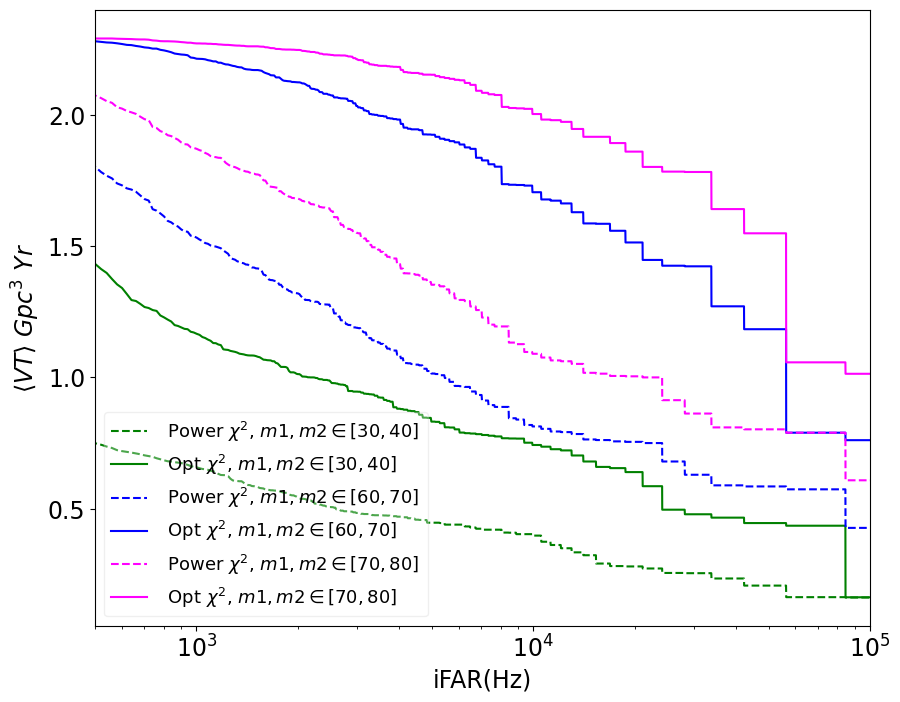}

\caption{ Here we show how the optimised SG $\chi^2$ compares with power $\chi^2$ in three different mass bins in terms of volume-time sensitivity. The component mass values are in units of $M_\odot$.
}
\label{vt_2}
\end{figure}

To compute the matched-filtering SNRs of both blips and spin-aligned BBH signals, we use different CBC template banks for the two mass bins. For the lower mass bin, $m_1,\>m_2\in[20,40]$, we use a template bank with component masses $(m_1,m_2)_{\rm bank} \in [10, 50] M_{\odot}$ and for the higher mass bin, $m_1,\>m_2\in[60,80] M_{\odot}$, we use a template bank with component masses $ (m_1,m_2)_{\rm bank}  \in [50, 90] M_{\odot}$. The template banks are intentionally chosen to cover a broader range of masses than the injections. For both mass bins, the templates had the same range of spins as the injections, namely,   $s_{1z}, s_{2z} \in [0.0, 0.9]$. 
Templates banks were generated with a minimum mismatch of 0.97 using the stochastic bank code of \textsc{PyCBC} \cite{sbank_2008}. We used the same \textsc{IMRPhenomPv2} waveform for templates as was used for simulating the CBC injections.
The BBH signals so prepared are injected into 16 sec long data segments from O3 that are not known to have any astrophysical signals in them. These 
data segments are multiplied with the Tukey window to make a smooth transition to zero strain at the edges, which mitigates spectral leakage during the analysis. 

We use the receiver-operating characteristic (ROC) curves to assess the performance of the optimal $\chi^2$ and compare it with that of the traditional $\chi^2$ and sine-Gaussian $\chi^2$ statistics. First, the optimal, traditional and sine-Gaussian $\chi^2$s are computed for the chosen sample of BBHs and blips, in addition to their respective SNRs.  We then use the ranking statistic
defined in Ref.~\cite{pycbc_soft, 2016_pycbc_usman} to rank the triggers in case of traditional $\chi^2$. 
On the other hand, for sine-Gaussian and optimal $\chi^2$ we use the new ranking statistic
defined in Ref.~\cite{Nitz:2017lco}. The ranking statistics used here for the traditional $\chi^2$ and the sine-Gaussian $\chi^2$ are the usual ones, respectively. The decision to use the new ranking statistic for the optimized SG $\chi^2$ was made after finding how both ranking statistics performed for the optimal $\chi^2$. As one can see from the ROC curves in \Fref{scatter_1}, 
{\it prima facie}, for these choices of the ranking statistics the optimized SG $\chi^2$ appears to perform better than the traditional and sine-Gaussian $\chi^2$ at all false alarm probabilities in both mass bins. 
Specifically, the optimized SG $\chi^2$ 
achieves an improvement in sensitivity of around 4\% over the traditional $\chi^2$ at a false alarm probability of $10^{-2}$.
For higher masses  $m_1,m_2 \in [60,80] M_{\odot}$ too we see that the optimized SG $\chi^2$ shows notable improvement in at and around the same false alarm probability.

Here, it is important to note that one is not discounting the possibility that further tuning of the Power and sine-Gaussian $\chi^2$s on the same injections and glitches would improve their performance. In fact, we did not exhaustively  tune the optimized SG $\chi^2$ either.

To check if the aforementioned performance improvement is limited to only a specific SNR bin, we repeated the above study for somewhat higher SNRs, namely SNR$~\in [8,12]$  in \Fref{scatter_2} (as compared to SNR$~\in [8,12]$ in \Fref{scatter_1}).
To further test if the improvement of the new statistic was limited in source distance, 
we analysed its performance in a couple of broad distance bins in
Fig.~\ref{dist_1}. With the aforementioned caveat about tuning, we observe that the optimized SG $\chi^2$ is found to improve the sensitivity in each of these cases, although in the more distant bin the improvement tends to vanish at low false-alarm probabilities, as expected: Weak signals are noise dominated and are difficult to discern from noise transient by any statistic. 
We also compare the  volume-time sensitivities~\cite{Tiwari_2018, Abbott:2016nhf,2016ApJS} as functions of the inverse false-alarm rate 
for three different $\chi^2$ statistics in Figs.~\ref{vt_1} and \ref{vt_2}. 
For this study, we performed signal injections in O3 data, for various component mass ranges, up to a maximum distance of 6 Gpc, uniformly distributed within the comoving volume. The volume-time sensitivity is then calculated by dividing the number of recovered injections by the total number of injected signals, as detailed in~\cite{Tiwari_2018}.


As can be observed in these results, the optimized SG $\chi^2$ performs at least as good as or somewhat better than the other statistics. This is likely because  it 
accounts for both the differences and the similarities between blips and BBH waveforms quantitatively by utilizing the metric in the sine-Gaussian space~\cite{Joshi_2021}.
The performance improvement is superior for higher mass BBHs because those signals occupy a lower part of the frequency band compared to the blips. The improvement over the sine-Gaussian $\chi^2$ can be understood in terms of the choice of $Q$ and $f_0$ range while sampling the sine-Gaussian waveforms to construct the $\VG$  vectors and then subtracting the triggered templates from them.
This work makes a few advances compared to the previous implementation of the optimal $\chi^2$~\cite{Joshi_2021}.
For instance, here we  utilized the {\em complex} form of the sine-Gaussian waveforms to account for the phase of the signal. Moreover, the sine-Gaussian projection maps~\cite{sunil_2022} helped us in identifying appropriate regions in the sine-Gaussian parameter space to choose 
the $\VG$ vectors from for the construction of the optimized SG $\chi^2$ statistic.
These advances 
also helped in bringing about appreciable improvement in CBC search sensitivity over traditional and sine-Gaussian $\chi^2$.


\begin{figure}
\includegraphics[width=\columnwidth]{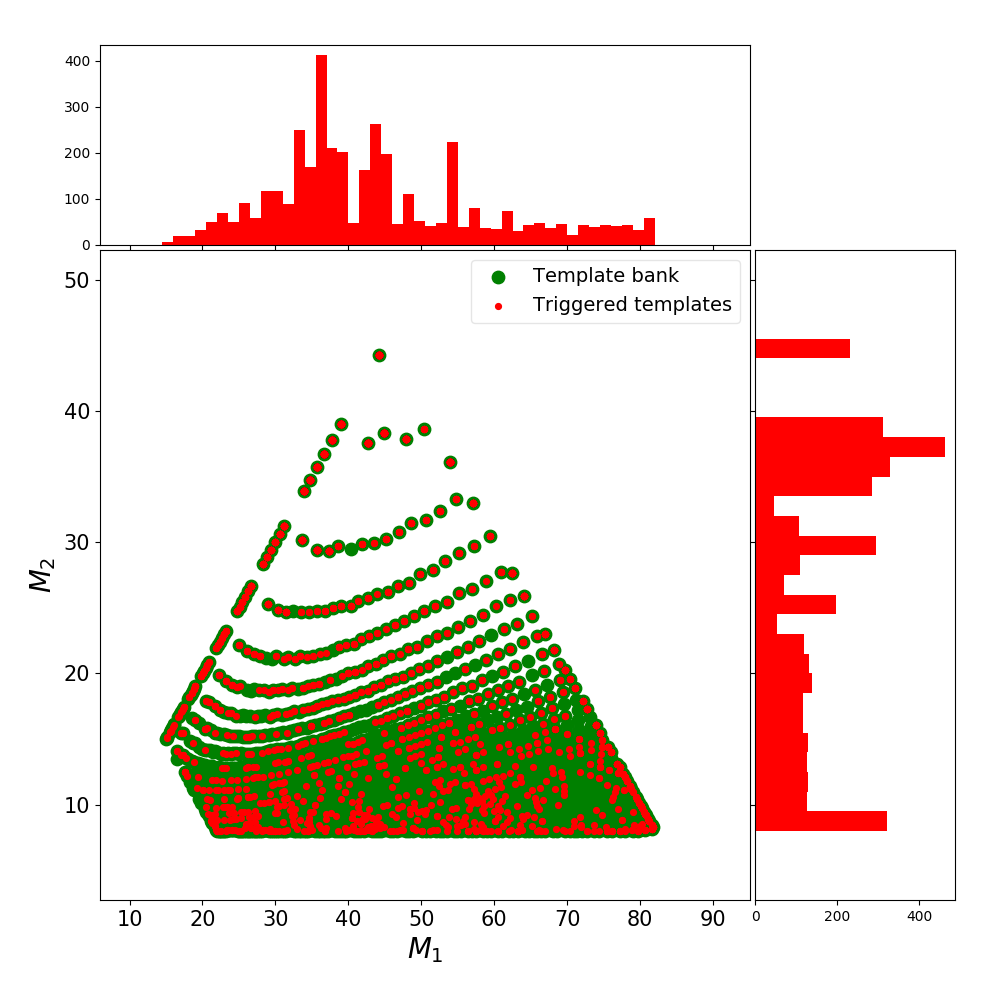}
\caption{The component masses of all templates employed are shown in green.
In contrast, the red ones (which are their subset) show the templates that are triggered by any of the (real) blips from a sample of 4000 -- taken from Livingston and Hanford detectors' O3 run. Corresponding to each mass axis, there is a histogram showing the number of templates triggered at each component mass value.}
\label{bank}
\end{figure}

It must be emphasized that just like for the traditional $\chi^2$, the computation of the optimized SG $\chi^2$ 
uses the parameter values of the triggered template. Therefore, the choice of template-bank boundaries while doing the matched-filtering operation can play an important role in its effectiveness. As we can see in the \Fref{bank}, blips can trigger a wide variety of templates, and a narrow template bank choice can adversely affect the performance of the optimized SG $\chi^2$.

\subsection{Performance on other noise transients}

\begin{figure*}
\includegraphics[scale=0.40]{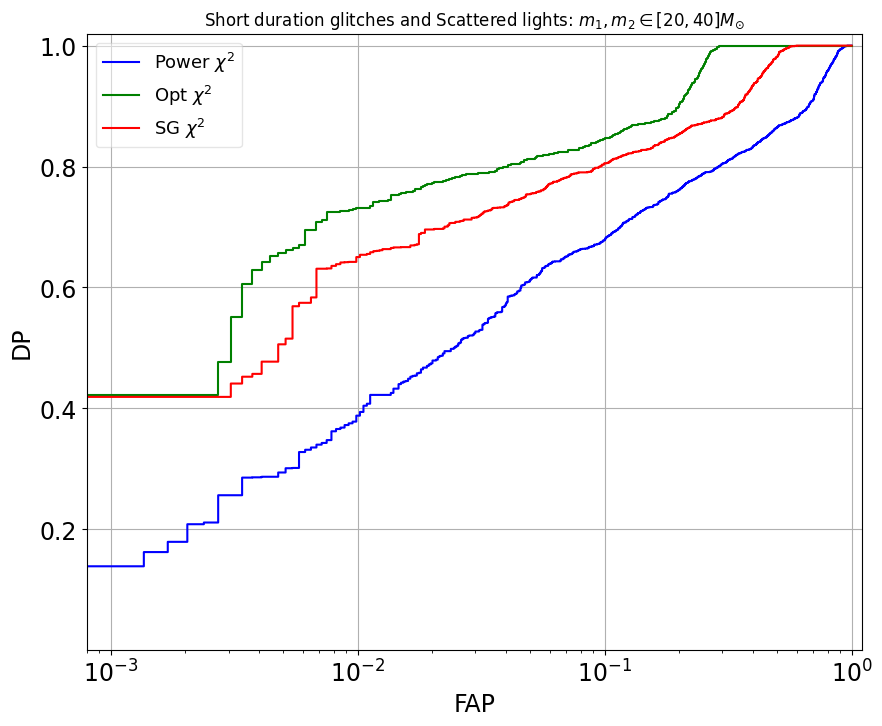}
\includegraphics[scale=0.40]{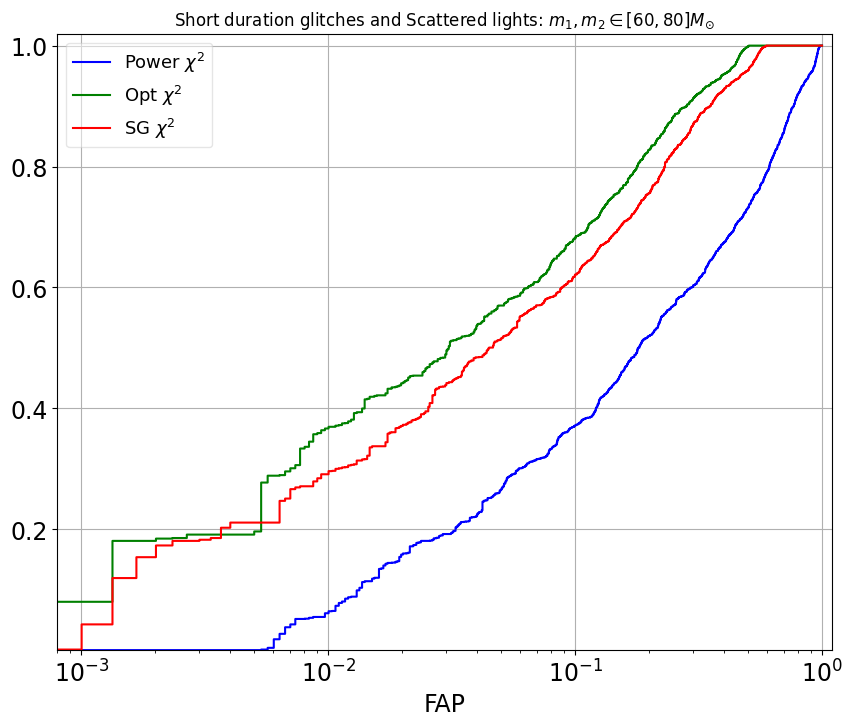}

\caption{ROC comparison between three $\chi^2$s for a set of short duration (koi fish, tomte, blip low-frequency blip) and scattered light glitches mixed against the simulated CBC signals in two component mass bins [20,40] $M_{\odot}$ (left) and [60,80] $M_{\odot}$ (right). Here the SNRs for CBC samples are uniformly distributed in the range [4,12]. }
\label{ROC_mixed}
\end{figure*}

\begin{figure*}
\includegraphics[scale=0.40]{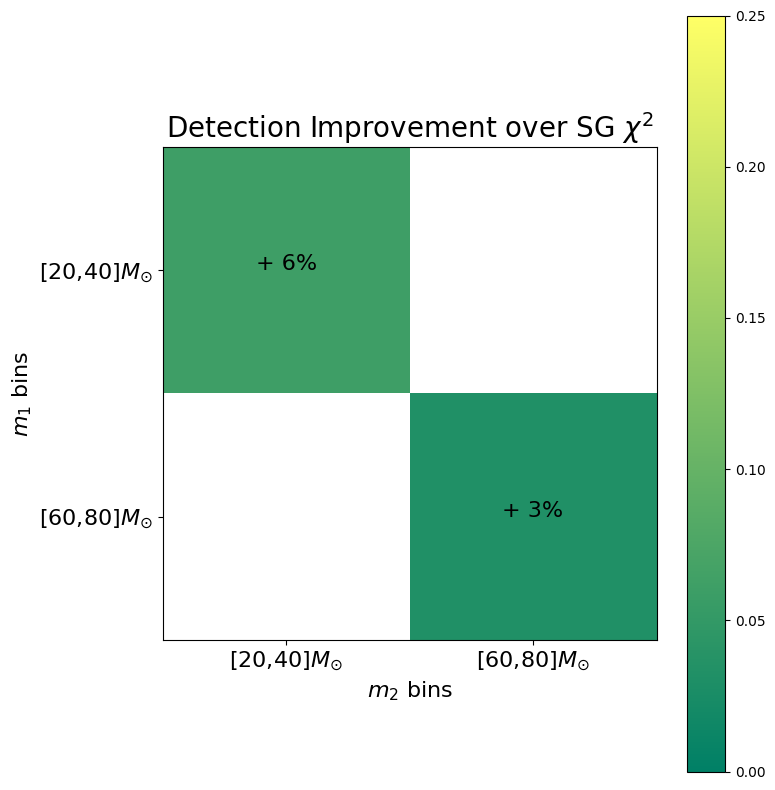}
\includegraphics[scale=0.40]{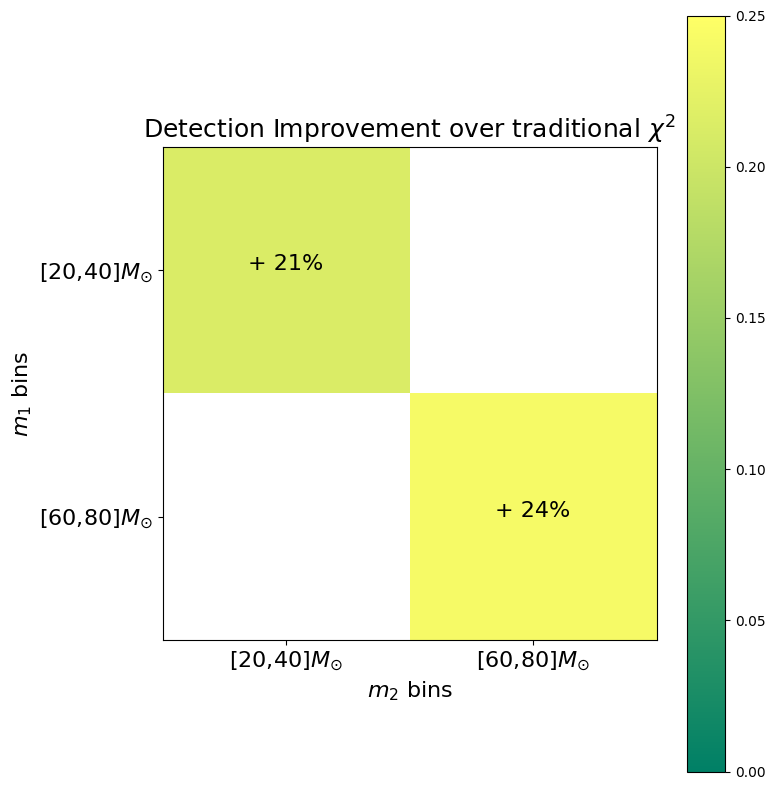}

\caption{Here we show the improvement in detection probability (true positive rate) of optimized SG $\chi^2$ over SG $\chi^2$ (left) and over traditional or power $\chi^2$ (right). The glitch sample in this case contain a mixture of 500 each of koifish, tomte, low-frequency blip, scattered light glitches and 1000 blips -- totalling 3000 glitches. Here the SNRs for the CBC samples are uniformly distributed in the range [4,12].
}
\label{dp1}
\end{figure*}

{So far we focused on applying the optimized SG $\chi^2$ for distinguishing BBH signals from the blip glitch and demonstrated
its utility via ROC curves.
In realistic observation scenarios, however, noise transients that trigger BBH templates are of a wider variety, and not limited to blips. To test the performance of the optimized SG $\chi^2$ on such a class, 
we compared the same three $\chi^2$s statistics as before on a set of mixed glitches, which include 500 each of koifish, tomte, low-frequency blip, scattered light glitches, in addition to 1000 blips --
which are known to occur more frequently. The performance comparison is shown as ROC plots in \Fref{ROC_mixed}. We observe that even in the presence of other kinds of noise transients in significant numbers, the optimized SG $\chi^2$ does slightly better than other two $\chi^2$s overall. \Fref{dp1} shows the improvement in the true positive rate by optimized SG $\chi^2$ over SG $\chi^2$ and traditional $\chi^2$. It is important to mention here that this improvement of optimized SG $\chi^2$ over other two $\chi^2$s is found without any tuning specific to these additional glitches.  To construct the basis vectors in this study, the sine-Gaussian waveforms were chosen from the region, $Q\in [2,8]$ and $f_0\in [100,500]$, as was done in the previous sections for blips.}

\section{Discussion}
\label{discussion}

Owing to their time-frequency morphological similarity with high-mass BBH signals, the short-duration noise transients like blips affect the search sensitivity of those signals adversely. In this study, we showed how the optimized SG $\chi^2$ can be constructed in real LIGO data so as to reduce that impact. The previous version of this $\chi^2$ statistic, introduced in 
Ref.~\cite{Joshi_2021}, is extended here to better discriminate blip glitches in searches for spin-aligned BBH signals in real data. 
A few advances have been made here 
compared to the past work in Ref.~\cite{Joshi_2021} that possibly helped in achieving further 
improvement. The first
one was accounting for the phase of the signal in the construction of the basis vectors -- by using complex sine-Gaussian waveforms. 
The second 
contributor 
was the use of projection maps~\cite{sunil_2022}, which 
guide us in locating the 
region in 
sine-Gaussian parameter space from which the initial sine-Gaussians should be chosen for constructing the basis vectors used in computing the optimized SG $\chi^2$.

The computational cost per trigger of the optimized SG $\chi^2$ is higher than that of the traditional $\chi^2$ and the sine-Gaussian $\chi^2$. A major fraction of this cost arises from the construction of orthonormal basis vectors described in~\Sref{SVD}. 
This makes the implementation of the optimized SG $\chi^2$ in an online search
less efficient computationally.
A straightforward way to 
reduce this inefficiency is to construct the orthonormal basis vectors  beforehand for a set of template masses.
In case of a high-mass template-bank, this advance preparation can be done for all the templates, as their number is relatively small. In case of low-mass template-banks, a more sparsely-populated fraction of templates can be selected for pre-computation of the orthonormal basis vectors. The optimized SG $\chi^2$ is then computed using the basis vectors of the nearest template. 


As noted before, there have been some attempts to veto blip glitches from the data with the help of a $\chi^2$ like statistic~\citep{Nitz:2017lco} and machine learning networks~\citep{connor_2022, sunil_2022} -- claiming varying degrees
of improvement in the BBH search sensitivity. Some of these works exploit certain insights provided by the \emph{unified} $\chi^2$ formalism~\citep{Dhurandhar_2017} but so far none of them fully 
leverages the power afforded by it.
The work reported in this paper attempted to bridge that gap -- for non-spinning as well as spin-aligned BBH signals. One could possibly extend this work to explore possible mitigation of the blips' effect on more general spinning CBC searches.



\section{Acknowledgment}

We would like to thank Sudhagar Suyamprakasam and Tanmaya Mishra for discussions related to LIGO data. 
We also thank Bhooshan Gadre for reviewing the manuscript and sharing some useful comments.
SVD acknowledges the support of the Senior Scientist Platinum Jubilee Fellowship from the National Academy of Sciences, India (NASI). 
The data analysis and simulations for this work were carried out at the IUCAA computing facility Sarathi. 
This research has made use of data, software, and/or web tools obtained from the Gravitational Wave Open Science Center \citep{gw_open_data, gwosc}, a service of LIGO Laboratory, the LIGO Scientific Collaboration, and the Virgo Collaboration. 
Some of that material and LIGO are
funded by the National Science Foundation. Virgo is funded by the French Centre National de Recherche Scientifique (CNRS), the Italian Istituto Nazionale della Fisica Nucleare (INFN), and the Dutch Nikhef, with contributions by Polish and Hungarian institutes. 
We also use some of the modules from the PyCBC an open source software package \citep{pycbc_soft}. This work was funded in part by NSF under Grant PHY-2309352.

\bibliography{Op_chi.bib}

\end{document}